# Anisotropy and Isotope Effect in Superconducting Solid Hydrogen

Mehmet Dogan[1,2,3], James R. Chelikowsky[1,4,5], Marvin L. Cohen[2,3*]

[1] Center for Computational Materials, Oden Institute for Computational Engineering and Sciences, University of Texas at Austin, Texas 78712, USA

[2] Department of Physics, University of California, Berkeley, CA 94720, USA

[3] Materials Sciences Division, Lawrence Berkeley National Laboratory, Berkeley, CA 94720, USA

[4] McKetta Department of Chemical Engineering, University of Texas at Austin, Texas 78712, USA

[5] Department of Physics, University of Texas at Austin, Texas 78712, USA

[*] To whom correspondence should be addressed: mlcohen@berkeley.edu

**Abstract**

Elucidating the phase diagram of solid hydrogen is a key objective in condensed matter physics. Several decades ago, it was proposed that at low temperatures and high pressures, solid hydrogen would be a metal with a high superconducting transition temperature. This transition to a metallic state can happen through the closing of the energy gap in the molecular solid or through a transition to an atomic solid. Recent experiments have managed to reach pressures in the range of 400–500 GPa, providing valuable insights. There is strong evidence suggesting that metallization via either of these mechanisms occurs within this pressure range. Computational and experimental studies have identified multiple promising crystal phases, but the limited accuracy of calculations and the limited capabilities of experiments prevent us from determining unequivocally the observed phase or phases. Therefore, it is crucial to investigate the superconducting properties of all the candidate phases. Recently, we reported the superconducting properties of the $C2/c$-24, $Cmca$-12, $Cmca$-4 and $I4_1/amd$-2 phases, including anharmonic effects. Here, we report the effects of anisotropy on superconducting properties using Eliashberg theory. Then, we investigate the superconducting properties of deuterium and estimate the size of the isotope effect for each phase. We find that the isotope effect on superconductivity is diminished by anharmonicity in the $C2/c$-24 and $Cmca$-12 phases and enlarged in the $Cmca$-4 and $I4_1/amd$-2 phases. Our anharmonic calculations of the $C2/c$-24 phase of deuterium agree closely with the most recent experiment by Loubeyre *et al.* [Phys. Rev. Lett. **29**, 035501 (2022)], indicating that the $C2/c$-24 phase remains the leading candidate in this pressure range, and has a strong anharmonic character. These characteristics can serve to distinguish among crystal phases in experiment. Furthermore, expanding our understanding of superconductivity in pure hydrogen holds significance in the study of high-$T_c$ hydrides.

At low temperatures, hydrogen takes the form of a molecular solid, which was initially theorized in 1935 to transform into a metallic atomic crystal [1]. Subsequently, following the development of the Bardeen–Cooper–Schrieffer (BCS) theory of superconductivity, it was suggested in 1968 that atomic solid hydrogen might exhibit a high superconducting transition temperature [2]. Nevertheless, even though hydrogen is the simplest atom and $H_2$ is the simplest molecule, the same simplicity does not extend to the crystal structure of solid hydrogen. Understanding the crystal structures of the molecular and atomic phases has proven to be a complex challenge both theoretically and experimentally. Additionally, predicting the pressure at which the transition to the atomic phase occurs is difficult, and only recently, after several decades of advancements in diamond anvil cell techniques, has experiment been able to begin to address such issues. To further complicate matters, a series of structural phase transitions may take place in solid hydrogen with increasing pressure, rather than a single transition from a molecular to an atomic phase.

Several potential candidates for the molecular and atomic phases were identified through theoretical investigations in the early 2000s [3–6]. Using sophisticated computational techniques enabling more precise enthalpy comparisons, along with insights from experimental data, the number of viable candidates has been reduced. Ultimately, three molecular phases (*C2/c*-24, *Cmca*-12, *Cmca*-4) and one atomic phase (*I4$_1$/amd*-2) have emerged as the most promising contenders [7–16]. (We use the notation where the number after the dash corresponds to the number of atoms in the unit cell.)

At moderate pressures, solid molecular hydrogen is a semiconductor. As pressure is increased, various scenarios may lead to metallization: (i) the same crystal phase could undergo a transition to a semimetallic and then to a metallic phase by closing the band gap, (ii) a structural transition from one molecular phase to another (semimetallic or metallic) could occur, or (iii) a direct structural transition to an atomic phase could take place. Computational studies are limited in conclusively predicting a particular scenario because of the presence of multiple crystal phases whose enthalpies lie within a few meV of each other at relevant pressures and the influence of factors related to the quantum nature of light hydrogen nuclei. However, computational investigations have successfully identified a few competitive phases within the pressure range of 300–500 GPa.

The primary experimental difficulty has been achieving higher pressures while ensuring the quality of the sample and facilitating accurate measurements. Over the course of the 2000s and 2010s, advancements were made, gradually extending the attainable pressure range to 400 GPa. This breakthrough led to the discovery of black hydrogen at 310–320 GPa, which exhibited no transmission in the visible range, indicating that the direct band gap is below ~1.5 eV [17–20], and later to the observation of heightened conductivity at 350–360 GPa, indicating semimetallic behavior [21–23]. Recently, a significant experiment conducted by Loubeyre et al. [24] has provided the most relevant data to date for determining the crystal structure within the pressure range of 150–425 GPa. The experiment employed infrared (IR) absorption measurements, which tracked the vibron frequency and the direct electronic band gap. The results indicated that the IR-active vibron frequency exhibits a linear decrease as pressure increases from 150 GPa to 425 GPa, suggesting the stability of a single phase within this pressure range. Furthermore, the direct band gap was observed to gradually decreases from 360 GPa to 420 GPa, but then suddenly drops below the experimentally observable threshold of ~0.1 eV. Although disagreements exist among the limited number of experimental groups engaged in high-pressure studies on solid hydrogen, some consensus has been reached, particularly regarding the heightened absorption in the range of 425–440 GPa, as found in the Loubeyre et al. study [24]. The experimental results that are concordant with these consist of the IR measurements conducted around 425 GPa by Dias et al. [21] and the Raman measurements around 440 GPa by Eremets et al. [22]. Little information has been reported regarding the behavior beyond 440 GPa. It is plausible, though, that a phase transition to atomic metallic hydrogen may occur around 500 GPa [25,26].

Previously, we reported a study investigating the changes in IR-active vibron frequencies within the *C2/c*-24 phase, using anharmonic corrections [27]. Our findings closely aligned with the observations made by Loubeyre et al. [24] up to 425 GPa. Additionally, we demonstrated that the observed variations in the direct band gap can be explained as modifications in the band structure as pressure increases, without necessitating a structural phase transition at 425 GPa. Consequently, a plausible scenario is that hydrogen may persist in the C2/c-24 phase beyond 425 GPa, potentially up to 500 GPa. However, it is also possible that a structural phase transition to a phase characterized by two vibron frequencies with a 300 cm$^{-1}$ difference occurs, as predicted by Dias *et al.* [21]. In this potential scenario, both the *Cmca*-12 and *Cmca*-4 phases would be candidates since they possess IR-active vibrons with approximately matching frequency differences [1]. Lastly, the I41/amd-2 atomic phase stands as a strong candidate for the atomic phase and might correspond to the observed crystal structure around 500 GPa [15,25,26]. As a result of this experimental and theoretical landscape, we include these four phases in our studies, which also aligns with the recent state-of-the-art attempts at computationally determining the phase diagram [28]. The superconducting properties of the phases with smaller unit cells (*Cmca*-4 and *I4$_1$/amd*-2) had been investigated computationally by other researchers [6,29–33], followed by our recent work on all the four phases [34,35], partly motivated by the rejuvenated interest in high-pressure superconducting hydrogen-rich systems [36–39].

In our previous treatment of the superconductivity of these phases of hydrogen, we employed the isotropic Eliashberg theory to estimate the superconducting temperatures of these systems [34,35]. Here, we expand our study to include anisotropic Eliashberg theory calculations. We also investigate the isotope effect by computing the vibrational properties,

electron–phonon coupling and superconducting properties of solid deuterium in the *C2/c*-24 *Cmca*-12, *Cmca*-4 and *I4₁/amd*-2 phases at 400, 450 and 500 GPa using density functional theory (DFT) calculations in the generalized gradient approximation [40–46], anharmonic corrections with the self-consistent phonon approach [47,48] and a Wannier function-based dense *k*-point and *q*-point sampling [49] (see the **Supplemental Material** for details).

The molecular phases *C2/c*-24, *Cmca*-12 and *Cmca*-4 consist of van der Waals-bonded layers of $H_2$ molecules, and the atomic phase consists of a highly symmetric 2-atom cell (the atomic structures can be found in the supplemental materials of refs. [34,35]). Among these four phases, as the number of atoms in the unit cell decreases, the structure becomes more symmetric, which is reflected in the electronic band structure. For the phases with larger unit cells, more complex Fermi surfaces exist with several sheets. The number of sheets also increases as a function of pressure. Given that the Allen–Dynes formula underestimates $T_c$ in this class of materials [50,51], it is important to conduct Eliashberg calculations to determine the leading edge of the superconducting gap ($\Delta_0$) *vs.* temperature, which we did previously using the isotropic approximation [34,35]. However, the complex nature of the Fermi surfaces in these phases necessitates further investigation into the effects of anisotropy. In **Figure 1(a),** we present the distribution of the superconducting gap on the Fermi surface for the *C2/c*-24 phase of hydrogen calculated in the anharmonic approximation at 500 GPa and 20 K. In this system, the five bands that cross the Fermi energy have $\Delta_0$ values in the same range (40–60 meV). **Figure 1(b)** shows the distribution of the superconducting gap for the same phase at 500 GPa, plotted for each computed temperature. The width of the distribution decreases as a function of temperature, as the average $\Delta_0$ approaches zero as expected. We find that the $T_c$ value of 242 K fits this behavior, and therefore the anisotropy does not modify $T_c$ [34]. We present the corresponding results for the harmonic calculation in **Figure S1**.

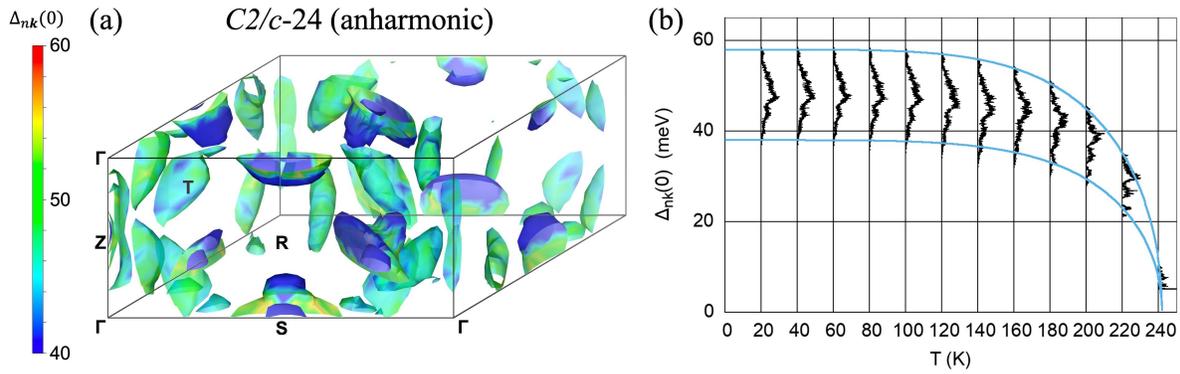

**Figure 1. Superconducting gap of the *C2/c*-24 phase at 500 GPa (anharmonic). (a)** The distribution of the superconducting gap on the Fermi surface (in meV) for the *C2/c*-24 phase of hydrogen calculated in the anharmonic approximation at 500 GPa and 20 K. **(b)** The distribution of the superconducting gap for the *C2/c*-24 phase of hydrogen calculated in the anharmonic approximation at 500 GPa, plotted for each computed temperature.

In **Figure 2(a),** we present the distribution of the superconducting gap on the Fermi surface for the *Cmca*-12 phase of hydrogen calculated in the anharmonic approximation at 500 GPa and 20 K. In this system, there are four bands that cross the Fermi energy, two of which have $\Delta_0$ values in the 25–35 meV range, and two of which have $\Delta_0$ values in the 35–55 meV range. **Figure 2(b)** shows the distribution of the superconducting gap for the same phase at 500 GPa, plotted for each computed temperature. The distributions do not include a gap, meaning that the sheets seen in **Figure 2(a)** have sufficiently overlapping distributions. Similar to the case of *C2/c*-24, the width of the distribution decreases as a function of temperature, as the average $\Delta_0$ approaches zero. We find that the $T_c$ value of 212 K fits this behavior, and therefore the anisotropy does not modify $T_c$ [35]. The corresponding results for the harmonic calculation are given in **Figure S2**.

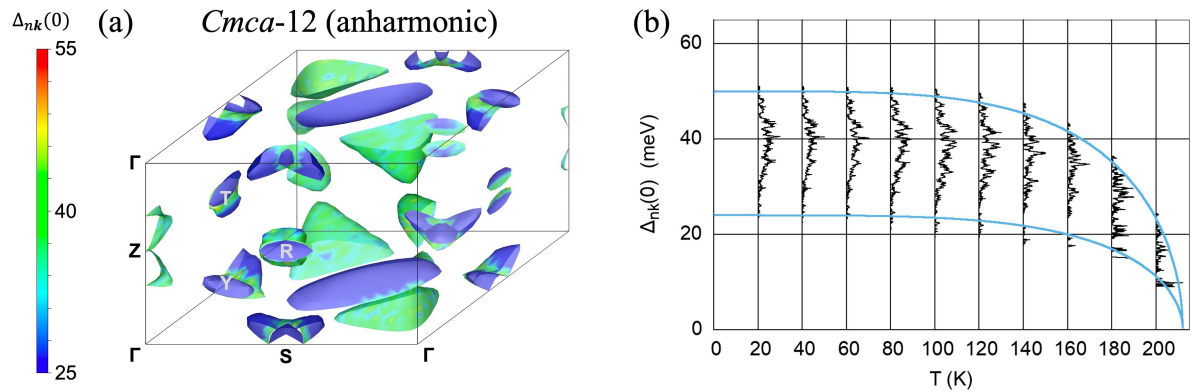

**Figure 2. Superconducting gap of the *Cmca*-12 phase at 500 GPa (anharmonic).** (a) The distribution of the superconducting gap on the Fermi surface (in meV) for the *Cmca*-12 phase of hydrogen calculated in the anharmonic approximation at 500 GPa and 20 K. (b) The distribution of the superconducting gap for the *Cmca*-12 phase of hydrogen calculated in the anharmonic approximation at 500 GPa, plotted for each computed temperature.

Next, we present the distribution of the superconducting gap on the Fermi surface for the *Cmca*-4 phase of hydrogen calculated in the anharmonic approximation at 500 GPa and 10 K in **Figure 3(a)**. In this system, there are two bands that cross the Fermi energy, which have $\Delta_0$ values in the 10–15 meV and 15–20 meV ranges, respectively. **Figure 3(b)** shows the distribution of the superconducting gap for the same phase at 500 GPa, plotted for each computed temperature. The distributions show a clear separation between the two sheets. These results are in agreement with a previous study that used superconducting density functional theory (SCDFT) [29,30]. The distribution of the superconducting gap on the Fermi surface for the *I4₁/amd*-2 phase of hydrogen calculated in the anharmonic approximation at 500 GPa and 20 K is presented in **Figure 4(a)**, whereas **Figure 4(b)** shows the distribution of the superconducting gap for the same phase at 500 GPa, plotted for each computed temperature. The four separate regions that are seen in **Figure 4(b)** correspond to the blue, cyan, green and orange parts of the sheets in **Figure 4(a)**, and no clear separation between the different sheets occurs, as opposed to the well-known case of $MgB_2$ [52–54]. The corresponding results for the harmonic calculations can be found in **Figure S3** for *Cmca*-4 and **Figure S4** for *I4$_1$/amd*-2.

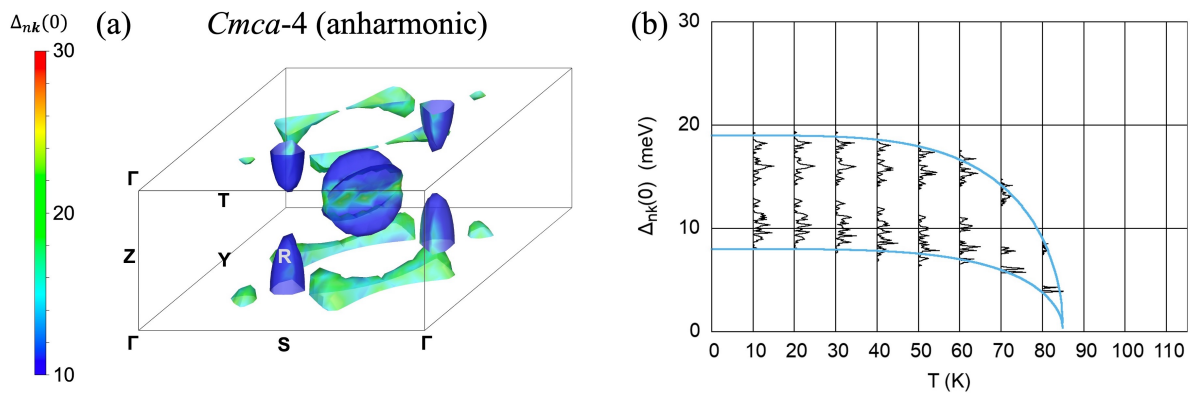

**Figure 3. Superconducting gap of the *Cmca*-4 phase at 500 GPa (anharmonic).** (a) The distribution of the superconducting gap on the Fermi surface (in meV) for the *Cmca*-4 phase of hydrogen calculated in the anharmonic approximation at 500 GPa and 10 K. (b) The distribution of the superconducting gap for the *Cmca*-4 phase of hydrogen calculated in the anharmonic approximation at 500 GPa, plotted for each computed temperature.

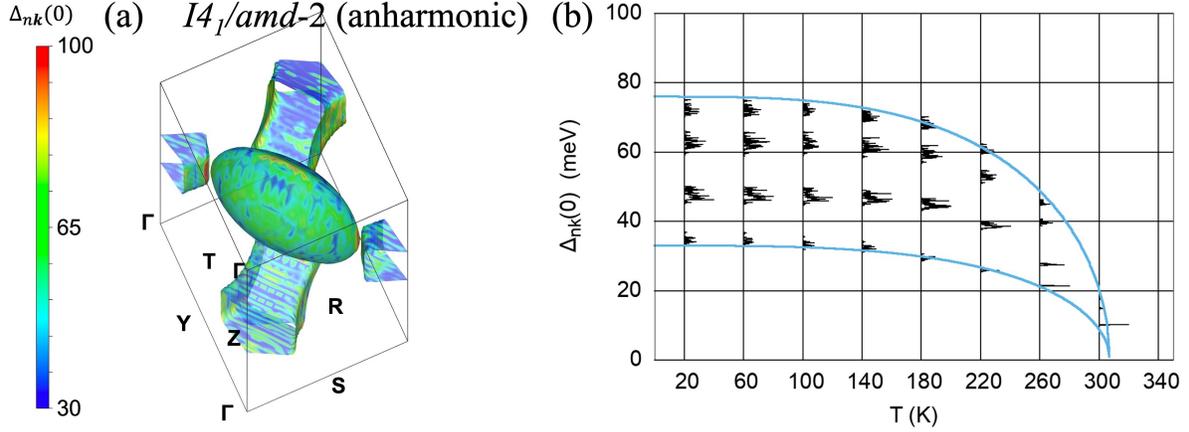

**Figure 4. Superconducting gap of the *I4₁/amd*-2 phase at 500 GPa (anharmonic).** (a) The distribution of the superconducting gap on the Fermi surface (in meV) for the *I4₁/amd*-2 phase of hydrogen calculated in the anharmonic approximation at 500 GPa and 20 K. (b) The distribution of the superconducting gap for the *I4₁/amd*-2 phase of hydrogen calculated in the anharmonic approximation at 500 GPa, plotted for each computed temperature.

Our investigation of the anisotropic Migdal–Eliashberg treatment of superconducting hydrogen has established that the electron–phonon superconductivity in these systems is not expected to be anomalous in the MgB$_2$ sense; i.e., the $T_c$ predictions based on the isotropic theory are expected to hold. However, the widened and/or multiple superconducting gaps should be measurable in experiment. We note here that our calculations are conducted as implemented in the EPW code, which assumes Lorentzian phonon lineshapes [49,55]. A recent preprint investigated the effects of non-Lorentzian phonon lineshapes in the context of the stochastic self-consistent harmonic approximation (SSCHA), and has found that they enhance anharmonicity and modify superconducting properties [56]. It is beyond the scope of this work to evaluate this proposed methodology; however, previous SSCHA calculations have overestimated band gaps and underestimated vibron frequencies for hydrogen [23,24,57], indicating that further investigations are needed. On the other hand, our simplified methodology has yielded close agreement for these quantities [27]. It is well known that band gaps are underestimated by the generalized gradient approximation (GGA). Utilizing more advanced techniques for considering excited states, such as the GW approximation, can increase band gaps by ~1.5 eV within this pressure range [7,10], and should have a notable effect on the *C2/c*-24 and *Cmca*-12 phases. Conversely, nuclear quantum effects have the opposite effect, causing a reduction in band gaps by smearing the position of potential wells. Interestingly, this effect occurs at a similar magnitude [58,59]. As a result of this fortuitous cancellation, we believe that the computed electronic structures provide a reliable approximation [27].

We now turn to our investigations of deuterium. In the harmonic approximation, doubling the size of the nuclei scales the phonon frequencies as well as $T_c$ by $1/\sqrt{2}$. However, our fully anharmonic calculations do not follow a simple rule, and provide an opportunity for a robust comparison with experiments. In **Figure S5, Figure S6,** and **Figure S7**, we present the phonon dispersion relations of the *C2/c*-24 phase at 400, 450 and 500 GPa, respectively, both in the harmonic and anharmonic approximations. To compare our results directly with experiment, we turn to recent work on deuterium by Loubeyre *et al.* [57], in which the frequency of the IR-active vibron was measured from 150 GPa up to 450 GPa, and follows a linear relationship with respect to pressure. The values for this frequency are 2890, 2830 and 2770 cm$^{-1}$ for 400, 450 and 500 GPa, respectively, where the last value is extrapolated [57]. Our anharmonic calculations yielded 2840, 2820 and 2790 cm$^{-1}$ for the same pressures. They also reported the frequency of the Raman-active vibron as a function of pressure up to 420 GPa. Extrapolating linearly, we get the values 2370, 2270 and 2170 cm$^{-1}$ for 400, 450 and 500 GPa, respectively. Our calculations for the Raman-active vibron frequencies yielded 2270, 2200 and 2060 cm$^{-1}$ for the same pressures.
As a measure of the isotope effect in the phonon spectrum, we can use $\beta$, which is defined by
$\omega(M_2) = \omega(M_1)\left(\frac{M_2}{M_1}\right)^{-\beta}$ and equal to 0.5 for the harmonic case. Using this definition and the
experiments on hydrogen and deuterium from the same group [24,57], we find $\beta$ to be 0.41, 0.39
and 0.37 for 400, 450 and 500 GPa, respectively, for the IR-active vibron. Using our previously
reported calculations on hydrogen [27,34] and the current ones on deuterium, we find $\beta$ to be
0.40, 0.39 and 0.38 for 400, 450 and 500 GPa, respectively. These results directly demonstrate
that the solid molecular hydrogen is distinctly anharmonic and can be described by the self-
consistent phonon approach used in our studies. We also note that the direct band gap of both
hydrogen and deuterium drop to zero from the same value as measured by Loubeyre *et
al.* [24,57], which would require a coincidence if this drop was caused by a structural phase
transition, but would be expected if it was caused by the changes in the band structure, as we
proposed previously [27]. For the *C2/c*-24 phase of deuterium, the Eliashberg function ($\alpha^2 F$),
the phonon densities of states (PhDOS), and the electron–phonon coupling parameter defined as
$\lambda(\omega) = \int_0^\omega \frac{d\omega'}{\omega'} \alpha^2 F(\omega')$ are presented in **Figure S5, Figure S6,** and **Figure S7** for 400, 450 and
500 GPa, respectively.

We have investigated the phonon dispersions, the Eliashberg function ($\alpha^2 F$), the phonon
densities of states (PhDOS), and the electron–phonon coupling parameter for the remaining
phases, reported in **Figure S8** through **Figure S16**. All of the calculated electron–phonon
coupling parameters ($\lambda$) and the superconducting transition temperatures ($T_c$) resulting from the
Allen–Dynes formula [60] ($\mu^* = 0.1$) as well as the isotropic Eliashberg formalism are reported
in **Table 1**. The $T_c$ values that resulted from the Eliashberg calculations are also shown in **Figure**

**5(a)**. Our results for deuterium show the same qualitative features as hydrogen which we have discussed previously [34,35]. For the purposes of this study, we are interested in the deviation of the isotope effect from the harmonic case. To that end, we plot $\beta$, defined (similarly to above) as $T_c(M_2) = T_c(M_1) \left(\frac{M_2}{M_1}\right)^{-\beta}$ in **Figure 5(b)** as a function of pressure for each phase. We find that the *C2/c*-24 and *Cmca*-12 phases have $\beta < 0.5$ whereas the *Cmca*-4 and $I4_1/amd$-2 phases have $\beta > 0.5$. In other words, for the former (latter) two phases, anharmonicity enhances (diminishes) superconductivity for the heavier isotope. This finding establishes that hydrogen is a complex system with significant anharmonicity and can be utilized to identify crystal phases in future experimental studies.

**Table 1. Electron–phonon coupling constant and $T_c$ of deuterium.** The electron–phonon coupling constant $\lambda$ and the superconducting transition temperature $T_c$ using both the Allen–Dynes formula and the isotropic Eliashberg theory are shown in the harmonic and anharmonic cases for 400, 450 and 500 GPa for the *C2/c*-24, *Cmca*-12, *Cmca*-4 and $I4_1/amd$-2 phases of deuterium. The Coulomb pseudopotential $\mu^*$ is set to 0.1 in all cases.

| Phase | P (GPa) | $\lambda$ | | $T_c$ (Allen–Dynes) (K) | | $T_c$ (Eliashberg) (K) | |
|---|---|---|---|---|---|---|---|
| | | harmonic | anharmonic | harmonic | anharmonic | harmonic | anharmonic |
| C2/c-24 | 400 | 0.43 | 0.54 | 6.0 | 1.1 | 6.1 | 1.2 |
| | 450 | 0.95 | 0.92 | 56 | 58 | 71 | 72 |
| | 500 | 1.80 | 1.48 | 131 | 148 | 173 | 185 |
| Cmca-12 | 400 | 0.87 | 0.94 | 57 | 54 | 68 | 63 |
| | 450 | 1.24 | 1.15 | 100 | 97 | 120 | 118 |
| | 500 | 1.49 | 1.27 | 123 | 128 | 149 | 158 |

| | 400 | 0.76 | 0.65 | 52 | 44 | 55 | 50 |
| Cmca-4 | 450 | 0.86 | 0.67 | 66 | 49 | 82 | 63 |
| | 500 | 0.89 | 0.64 | 68 | 52 | 78 | 58 |
| | 400 | 2.18 | 1.61 | 185 | 167 | 254 | 220 |
| $I4_1/amd$-2 | 450 | 2.16 | 1.33 | 184 | 156 | 259 | 227 |
| | 500 | 2.07 | 1.55 | 180 | 148 | 246 | 204 |

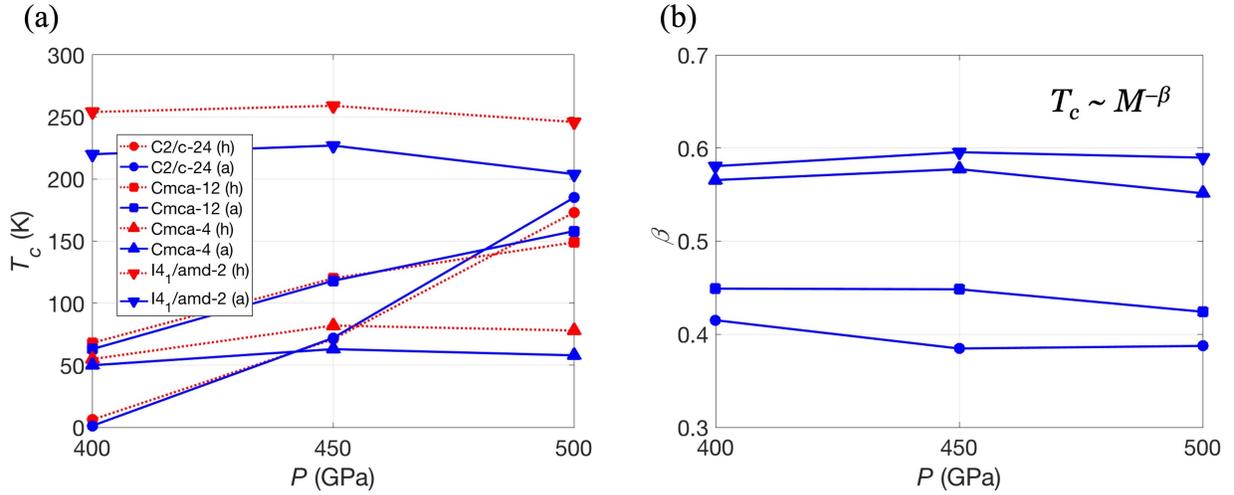

**Figure 5. $T_c$ for deuterium and the isotope effect. (a)** Superconducting transition temperature computed *via* the isotropic Eliashberg equations for the *C2/c*-24, *Cmca*-12, *Cmca*-4 and *I4₁/amd*-2 phases of deuterium. **(b)** The isotope effect for the *C2/c*-24, *Cmca*-12, *Cmca*-4 and *I4₁/amd*-2 phases of hydrogen, computed *via* the isotropic Eliashberg equations.

In summary, we demonstrated the effects of anisotropic Eliashberg theory on the superconducting properties of the *C2/c*-24, *Cmca*-12, *Cmca*-4 and *I4₁/amd*-2 phases of hydrogen.

We found that the anisotropy does not cause an anomalous change in the critical temperature, but it is significant and should be measurable. We also examined the superconducting properties of deuterium and estimated the extent of the isotope effect for each phase. Our findings reveal that anharmonicity diminishes the isotope effect in the *C2/c*-24 and *Cmca*-12 phases, while it is amplified in the *Cmca*-4 and *I4$_1$/amd*-2 phases. Importantly, we established agreement with the most recent experiment conducted by Loubeyre *et al.* [57] through our anharmonic calculations of the *C2/c*-24 phase of deuterium, which serves as evidence that the *C2/c*-24 phase remains the leading contender within this pressure range and its properties are well described by our theoretical approach. In future experiments, our findings can help differentiate between various crystal phases. Moreover, deepening our understanding of superconductivity in pure hydrogen has important implications for investigating high-temperature superconducting hydrides.

**Acknowledgements**

MLC acknowledges support by the Director, Office of Science, Office of Basic Energy Sciences, Materials Sciences and Engineering Division, of the U.S. Department of Energy under contract No. DE-AC02-05-CH11231, within the Theory of Materials program (KC2301), which supported the structure optimization and calculation of vibrational properties. MD acknowledges support from the "Characteristic Science Applications for the Leadership Class Computing Facility" project, which is supported by National Science Foundation Award No. 2139536. Computational resources used were Cori at National Energy Research Scientific Computing Center (NERSC), which is supported by the Office of Science of the US Department of Energy under contract no. DE-AC02-05CH11231, Stampede2 at the Texas Advanced Computing Center (TACC) through Extreme Science and Engineering Discovery Environment (XSEDE), which is supported by National Science Foundation (NSF) under grant no. ACI-1053575, Frontera at TACC, which is supported by NSF grant no. OAC-1818253, and Bridges-2 at the Pittsburgh Supercomputing Center (PSC), which is supported by NSF award number ACI-1928147. JRC acknowledges support by a subaward from the Center for Computational Study of Excited-State Phenomena in Energy Materials (C2SEPEM) at the Lawrence Berkeley National Laboratory, which is funded by the U.S. Department of Energy under contract no. DE-AC02- 05CH11231, as part of the Computational Materials Sciences Program. We thank Hyungjun Lee for technical assistance with the EPW code.

# Supplemental Material for "Anisotropy and Isotope Effect in Superconducting Solid Hydrogen"

Mehmet Dogan[1,2,3], James R. Chelikowsky[1,4,5], Marvin L. Cohen[2,3*]

[1] Center for Computational Materials, Oden Institute for Computational Engineering and Sciences, University of Texas at Austin, Texas 78712, USA

[2] Department of Physics, University of California, Berkeley, CA 94720, USA

[3] Materials Sciences Division, Lawrence Berkeley National Laboratory, Berkeley, CA 94720, USA

[4] McKetta Department of Chemical Engineering, University of Texas at Austin, Texas 78712, USA

[5] Department of Physics, University of Texas at Austin, Texas 78712, USA

[*] To whom correspondence should be addressed: mlcohen@berkeley.edu

**Computational Methods**

We compute optimized crystal structures using density functional theory (DFT) in the Perdew–Burke–Ernzerhof generalized gradient approximation (PBE GGA), [1] using the QUANTUM ESPRESSO software package. We use norm-conserving pseudopotentials with a 120 Ry plane-wave energy cutoff. [2,3] The Monkhorst–Pack $k$-point meshes we use to sample the Brillouin zone are listed in **Table S1**. All atomic coordinates are relaxed until the forces on all the atoms are less than $10^{-4}$ Ry/$a_0$ in all three Cartesian directions ($a_0$: Bohr radius). After each structure has been relaxed at a given pressure, a denser sampling of the Brillouin zone is examined to determine the band structure, using a k-point mesh that is twice as fine in all three reciprocal lattice directions. The Fermi surfaces are plotted using the XCrySDen package. [4]

Using density functional perturbation theory (DFPT), [5,6] we calculate the vibrational modes in the harmonic approximation for the pressures of 400, 450 and 500 GPa. Full phonon dispersions are calculated using the $q$-point samplings listed in **Table S1**. We apply the anharmonic corrections using the self-consistent phonon method as implemented in the ALAMODE package. [7,8] For the anharmonic corrections, we use the finite-displacement approach with the real-space supercells listed in **Table S1**. [9] The displacements from the equilibrium position are 0.01, 0.03, and 0.04 Å for the second, third, and fourth order terms of the interatomic force constant, respectively. We use the EPW code to implement the Wannier function-based $k$-space and $q$-space sampling for electron–phonon coupling calculations as well as isotropic and anisotropic Eliashberg theory calculations. [10] The Wannier interpolation is done using fine $k$-point and $q$-point samplings that are given in **Table S1**, using Wannierization with H 1s orbitals as initial projections, (–30 eV, 10 eV) as the inner window and (–30 eV, 70 eV) as the outer window around the Fermi energy.

**Table S1. Sampling parameters for all calculations.** All the sampling parameters for the calculations of the *C2/c*-24, *Cmca*-12, *Cmca*-4 and *I4₁/amd*-2 phases.

| Calculation stage | *C2/c*-24 | *Cmca*-12 | *Cmca*-4 | *I4$_1$/amd*-2 |
|---|---|---|---|---|
| *k*-point mesh for DFT | 16×16×8 | 16×16×16 | 32×32×16 | 32×32×32 |
| *q*-point mesh for DFPT | 4×4×2 | 4×4×4 | 8×8×4 | 8×8×8 |
| self-consistent phonon supercell | 2×2×1 | 2×2×2 | 4×4×2 | 4×4×4 |
| fine *k*-point mesh for e–ph coupling | 32×32×16 | 32×32×32 | 64×64×32 | 64×64×64 |
| fine *q*-point mesh for e–ph coupling | 8×8×4 | 8×8×8 | 16×16×8 | 16×16×16 |

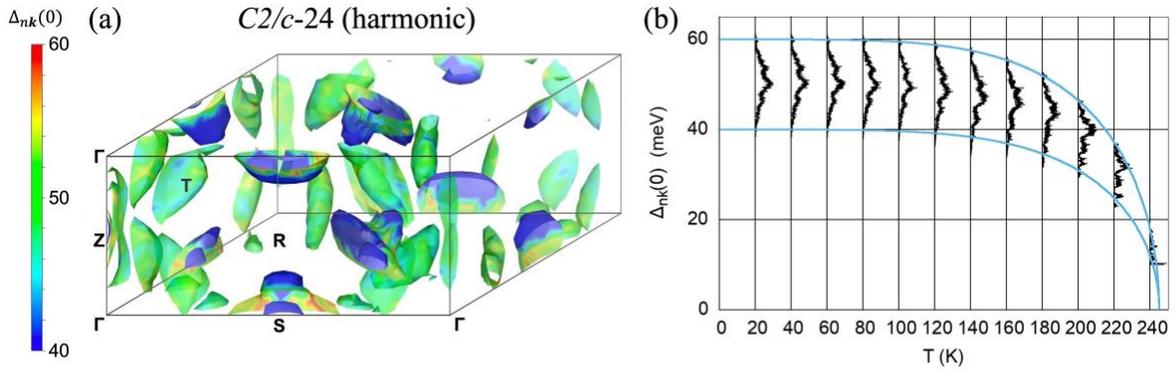

**Figure S1. Superconducting gap of the *C2/c*-24 phase at 500 GPa (harmonic).** (a) The distribution of the superconducting gap on the Fermi surface (in meV) for the *C2/c*-24 phase of hydrogen calculated in the harmonic approximation at 500 GPa and 20 K. (b) The distribution of the superconducting gap for the *C2/c*-24 phase of hydrogen calculated in the harmonic approximation at 500 GPa, plotted for each computed temperature.

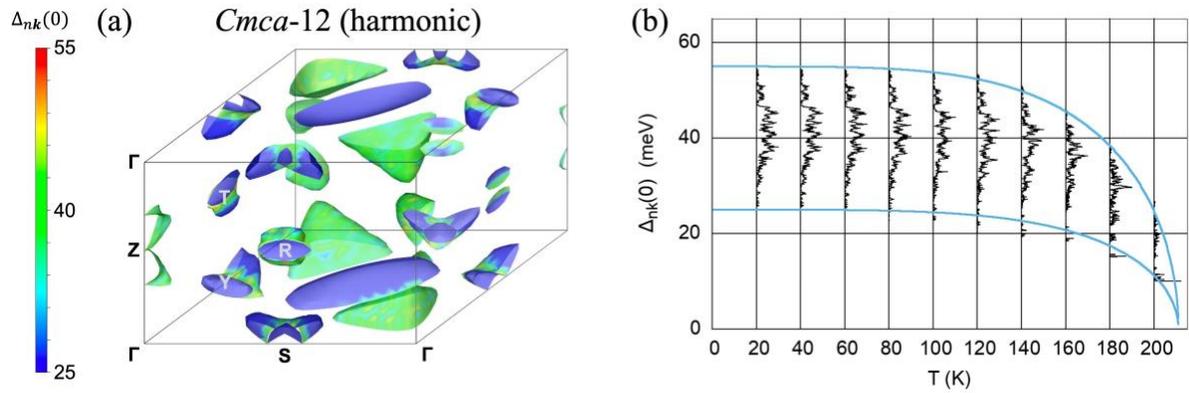

**Figure S2. Superconducting gap of the *Cmca*-12 phase at 500 GPa (harmonic).** (a) The distribution of the superconducting gap on the Fermi surface (in meV) for the *Cmca*-12 phase of hydrogen calculated in the harmonic approximation at 500 GPa and 20 K. (b) The distribution of the superconducting gap for the *Cmca*-12 phase of hydrogen calculated in the harmonic approximation at 500 GPa, plotted for each computed temperature.

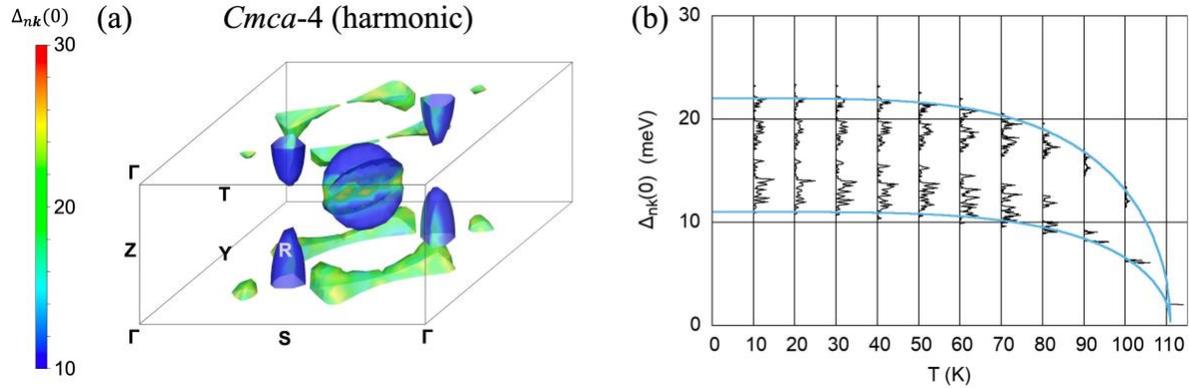

**Figure S3. Superconducting gap of the *Cmca*-4 phase at 500 GPa (harmonic).** (a) The distribution of the superconducting gap on the Fermi surface (in meV) for the *Cmca*-4 phase of hydrogen calculated in the harmonic approximation at 500 GPa and 10 K. (b) The distribution of the superconducting gap for the *Cmca*-4 phase of hydrogen calculated in the harmonic approximation at 500 GPa, plotted for each computed temperature.

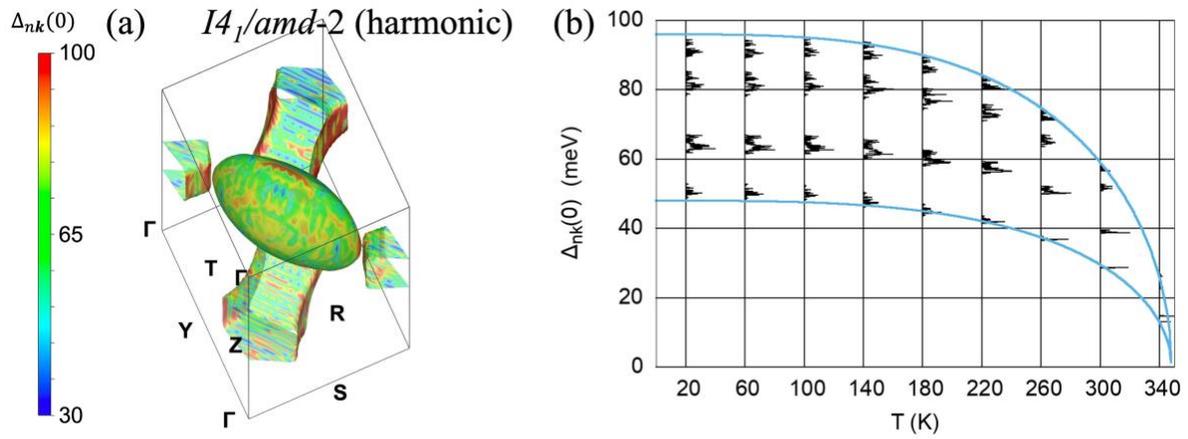

**Figure S4. Superconducting gap of the *I4$_1$/amd*-2 phase at 500 GPa (harmonic).** (a) The distribution of the superconducting gap on the Fermi surface (in meV) for the *I4$_1$/amd*-2 phase of hydrogen calculated in the harmonic approximation at 500 GPa and 20 K. (b) The distribution of the superconducting gap for the *I4$_1$/amd*-2 phase of hydrogen calculated in the harmonic approximation at 500 GPa, plotted for each computed temperature.

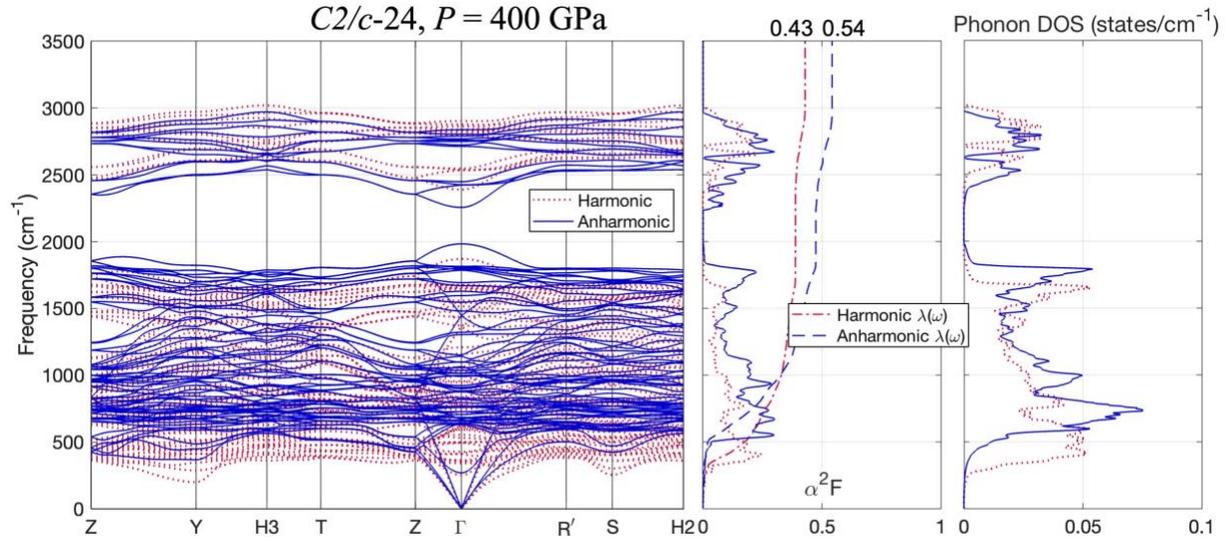

**Figure S5. Phonons and electron–phonon coupling for *C2/c*-24 at 400 GPa.** The phonon dispersion relations for the *C2/c*-24 phase of deuterium at 400 GPa pressure (left panel). The harmonic and anharmonic calculations are shown by red dashed lines and blue solid lines, respectively. The Eliashberg function $\alpha^2 F$ and the electron–phonon coupling parameter $\lambda(\omega)$ (middle panel), and the phonon densities of states (right panel).

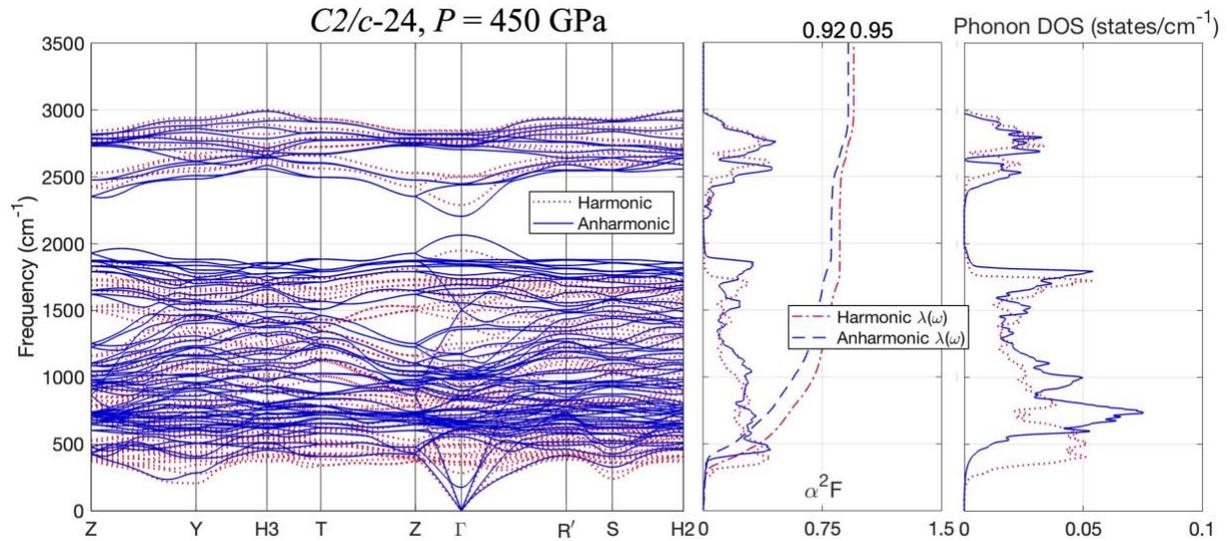

**Figure S6. Phonons and electron–phonon coupling for *C2/c*-24 at 450 GPa.** The phonon dispersion relations for the *C2/c*-24 phase of deuterium at 450 GPa pressure (left panel). The harmonic and anharmonic calculations are shown by red dashed lines and blue solid lines, respectively. The Eliashberg function $\alpha^2 F$ and the electron–phonon coupling parameter $\lambda(\omega)$ (middle panel), and the phonon densities of states (right panel).

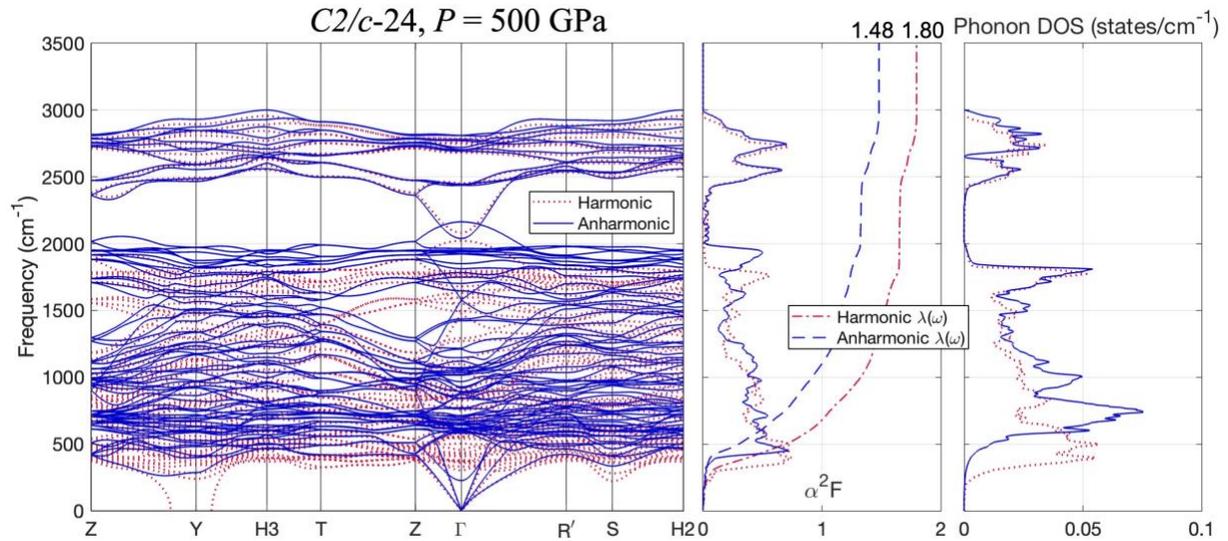

**Figure S7. Phonons and electron–phonon coupling for *C2/c*-24 at 500 GPa.** The phonon dispersion relations for the *C2/c*-24 phase of deuterium at 500 GPa pressure (left panel). The harmonic and anharmonic calculations are shown by red dashed lines and blue solid lines, respectively. The Eliashberg function $\alpha^2 F$ and the electron–phonon coupling parameter $\lambda(\omega)$ (middle panel), and the phonon densities of states (right panel).

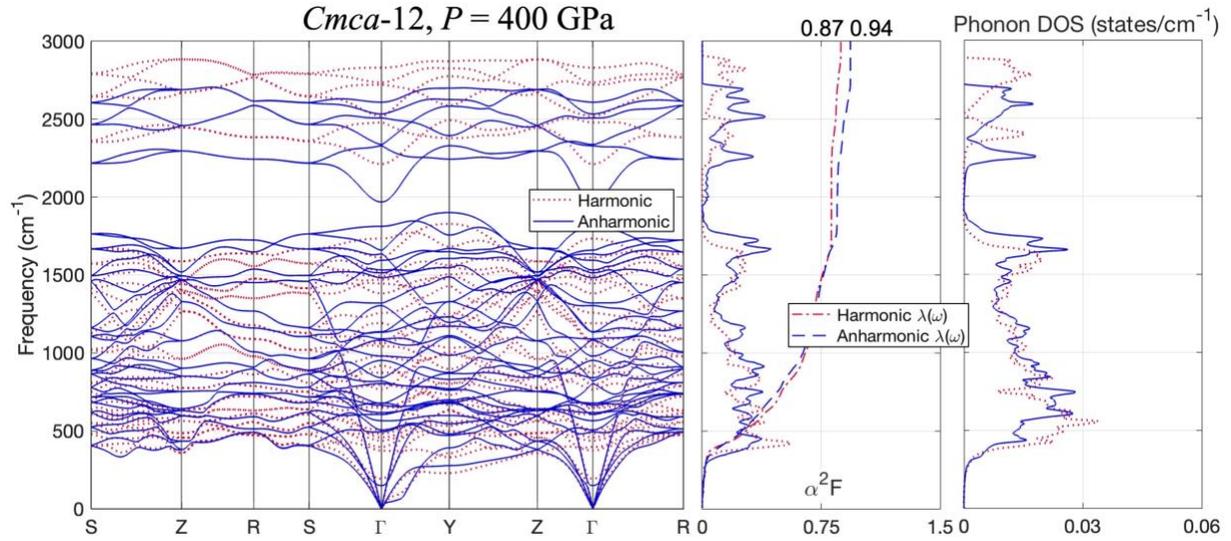

**Figure S8. Phonons and electron–phonon coupling for *Cmca*-12 at 400 GPa.** The phonon dispersion relations for the *Cmca*-12 phase of deuterium at 400 GPa pressure (left panel). The harmonic and anharmonic calculations are shown by red dashed lines and blue solid lines, respectively. The Eliashberg function $\alpha^2 F$ and the electron–phonon coupling parameter $\lambda(\omega)$ (middle panel), and the phonon densities of states (right panel).

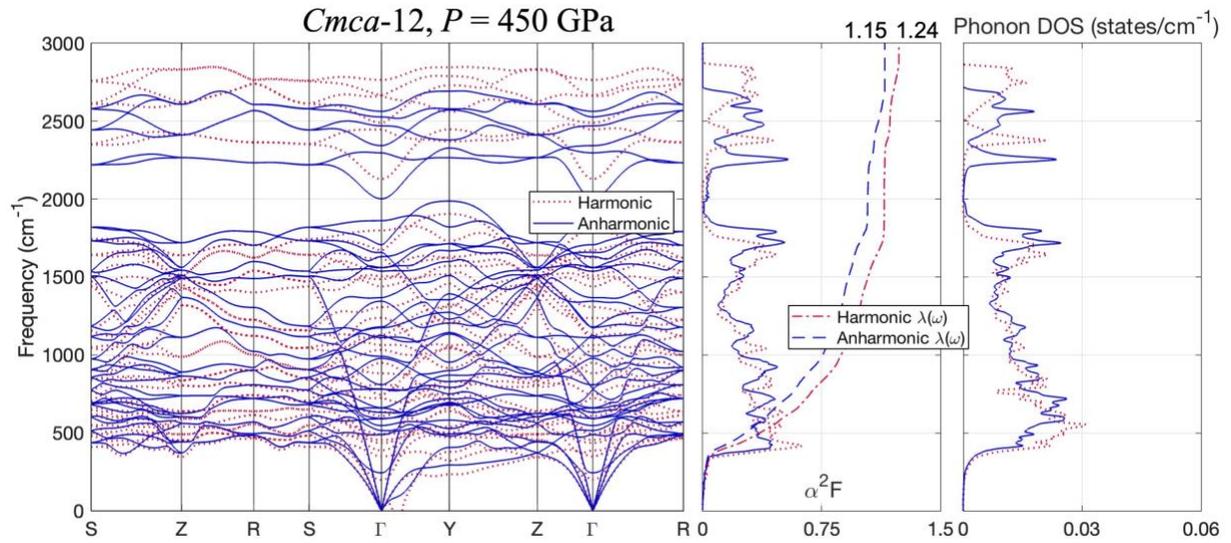

**Figure S9. Phonons and electron–phonon coupling for *Cmca*-12 at 450 GPa.** The phonon dispersion relations for the *Cmca*-12 phase of deuterium at 450 GPa pressure (left panel). The harmonic and anharmonic calculations are shown by red dashed lines and blue solid lines, respectively. The Eliashberg function $\alpha^2 F$ and the electron–phonon coupling parameter $\lambda(\omega)$ (middle panel), and the phonon densities of states (right panel).

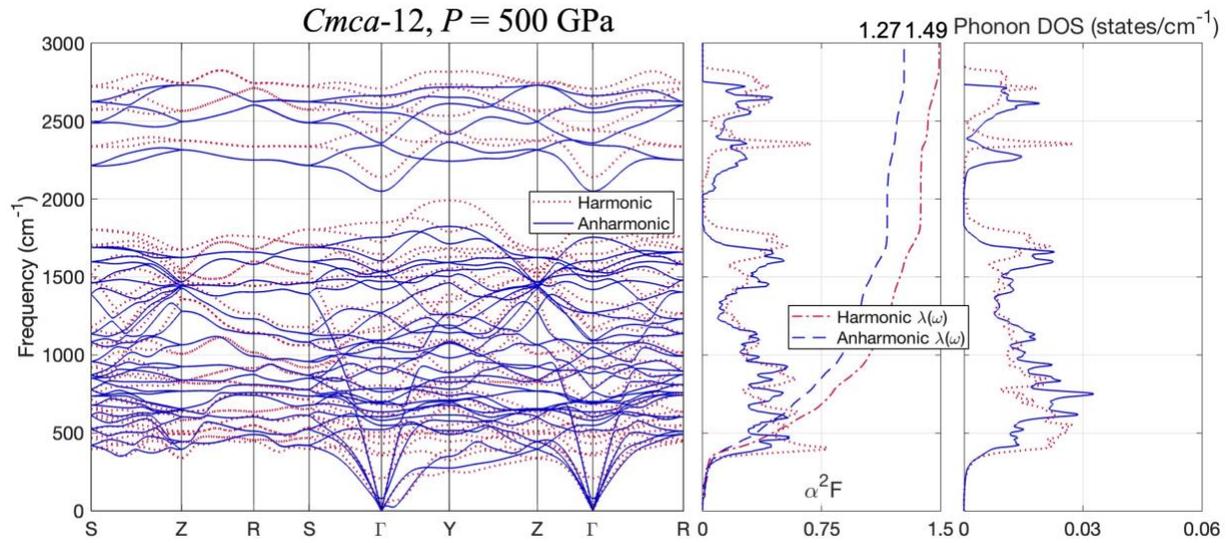

**Figure S10. Phonons and electron–phonon coupling for *Cmca*-12 at 500 GPa.** The phonon dispersion relations for the *Cmca*-12 phase of deuterium at 500 GPa pressure (left panel). The harmonic and anharmonic calculations are shown by red dashed lines and blue solid lines, respectively. The Eliashberg function $\alpha^2 F$ and the electron–phonon coupling parameter $\lambda(\omega)$ (middle panel), and the phonon densities of states (right panel).

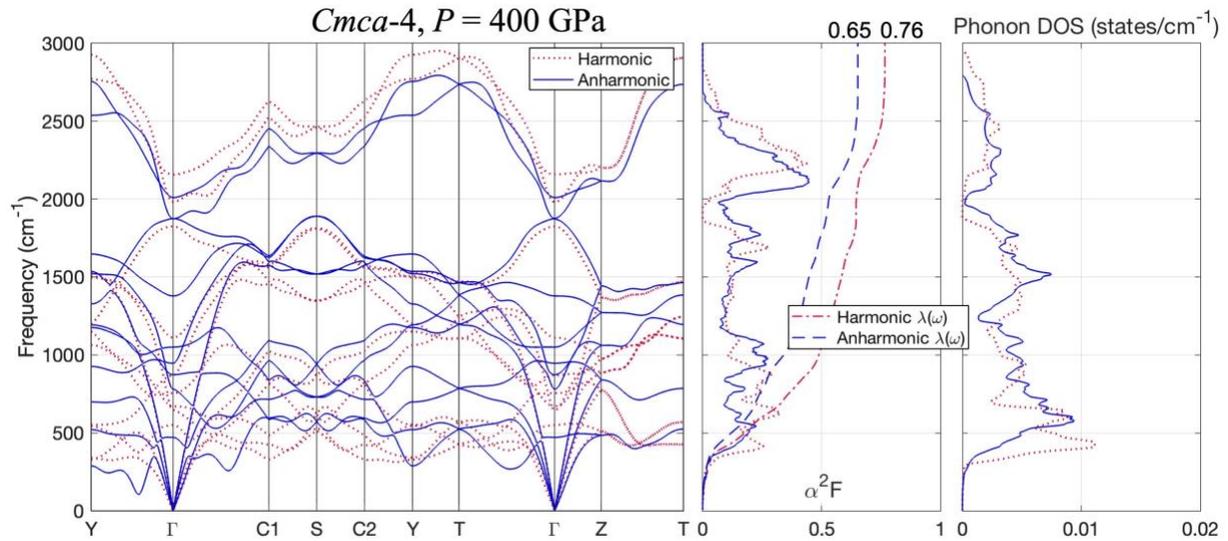

**Figure S11. Phonons and electron–phonon coupling for *Cmca*-4 at 400 GPa.** The phonon dispersion relations for the *Cmca*-4 phase of deuterium at 400 GPa pressure (left panel). The harmonic and anharmonic calculations are shown by red dashed lines and blue solid lines, respectively. The Eliashberg function $\alpha^2 F$ and the electron–phonon coupling parameter $\lambda(\omega)$ (middle panel), and the phonon densities of states (right panel).

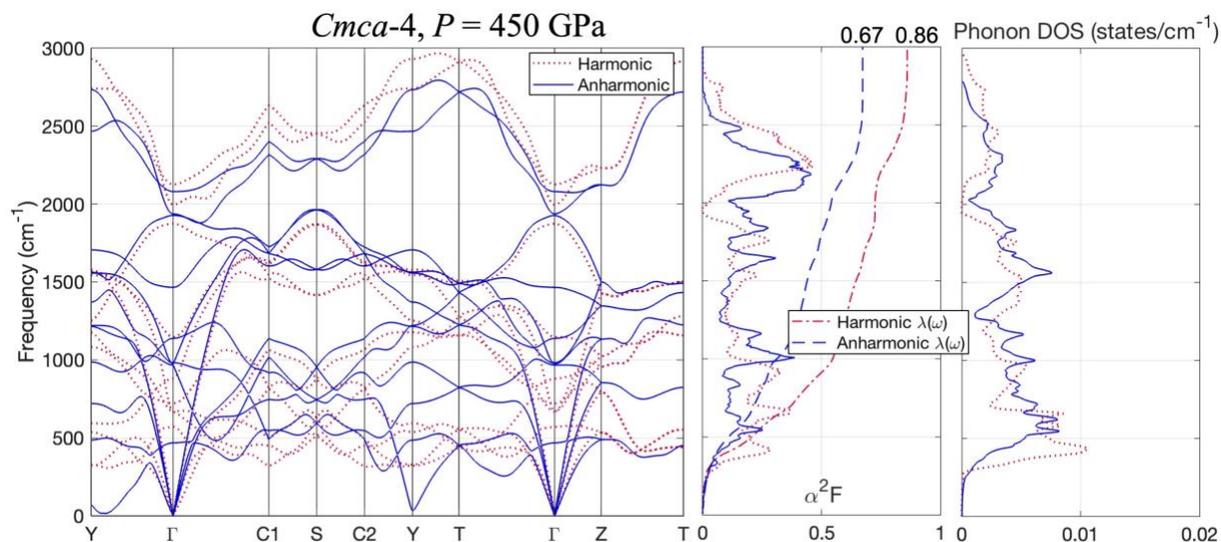

**Figure S12. Phonons and electron–phonon coupling for *Cmca*-4 at 450 GPa.** The phonon dispersion relations for the *Cmca*-4 phase of deuterium at 450 GPa pressure (left panel). The harmonic and anharmonic calculations are shown by red dashed lines and blue solid lines, respectively. The Eliashberg function $\alpha^2 F$ and the electron–phonon coupling parameter $\lambda(\omega)$ (middle panel), and the phonon densities of states (right panel).

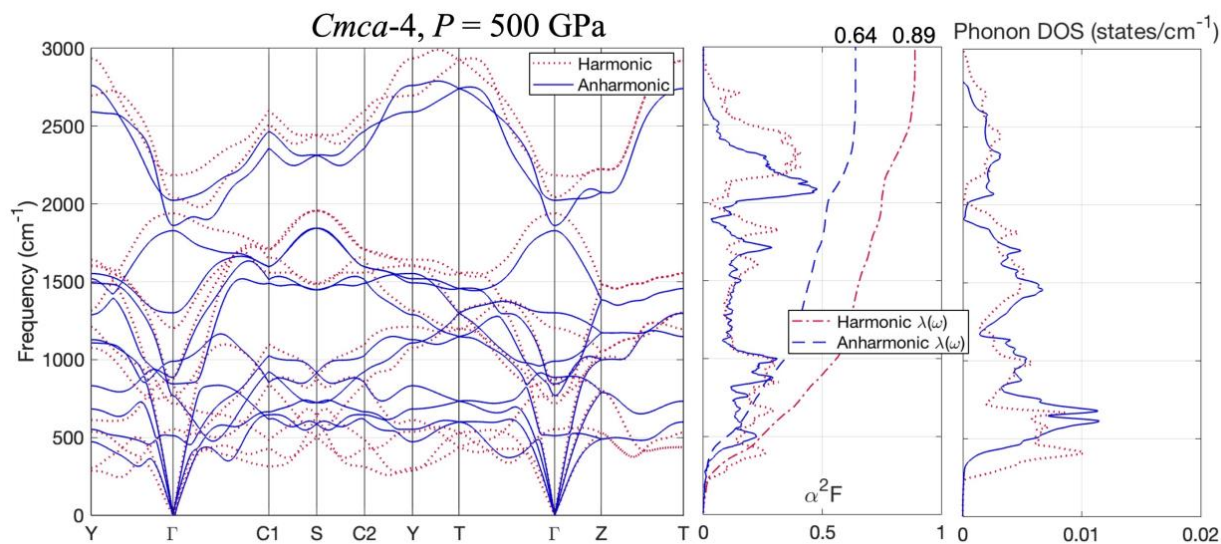

**Figure S13. Phonons and electron–phonon coupling for *Cmca*-4 at 500 GPa.** The phonon dispersion relations for the *Cmca*-4 phase of deuterium at 500 GPa pressure (left panel). The harmonic and anharmonic calculations are shown by red dashed lines and blue solid lines, respectively. The Eliashberg function $\alpha^2 F$ and the electron–phonon coupling parameter $\lambda(\omega)$ (middle panel), and the phonon densities of states (right panel).

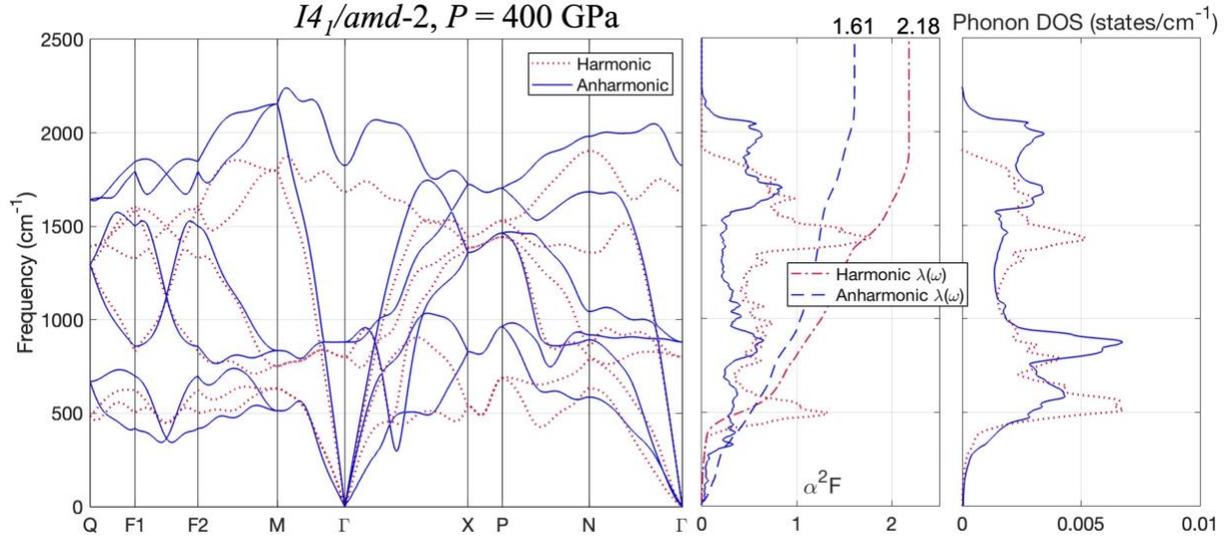

**Figure S14. Phonons and electron–phonon coupling for *I4₁/amd*-2 at 400 GPa.** The phonon dispersion relations for the *I4₁/amd*-2 phase of deuterium at 400 GPa pressure (left panel). The harmonic and anharmonic calculations are shown by red dashed lines and blue solid lines, respectively. The Eliashberg function $\alpha^2 F$ and the electron–phonon coupling parameter $\lambda(\omega)$ (middle panel), and the phonon densities of states (right panel).

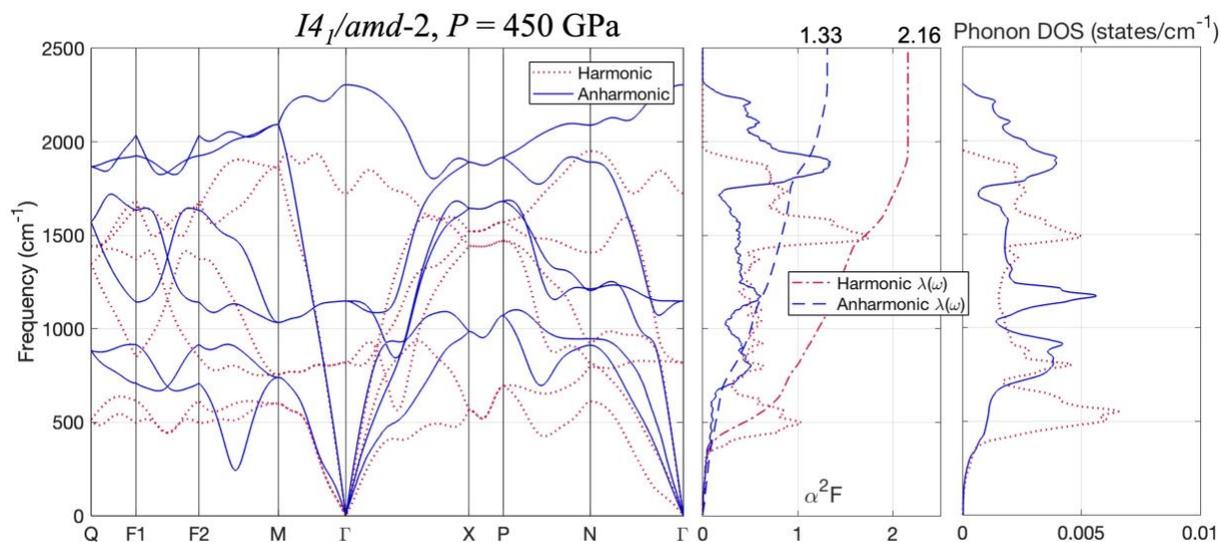

**Figure S15. Phonons and electron–phonon coupling for *I4₁/amd*-2 at 450 GPa.** The phonon dispersion relations for the *I4₁/amd*-2 phase of deuterium at 450 GPa pressure (left panel). The harmonic and anharmonic calculations are shown by red dashed lines and blue solid lines, respectively. The Eliashberg function $\alpha^2 F$ and the electron–phonon coupling parameter $\lambda(\omega)$ (middle panel), and the phonon densities of states (right panel).

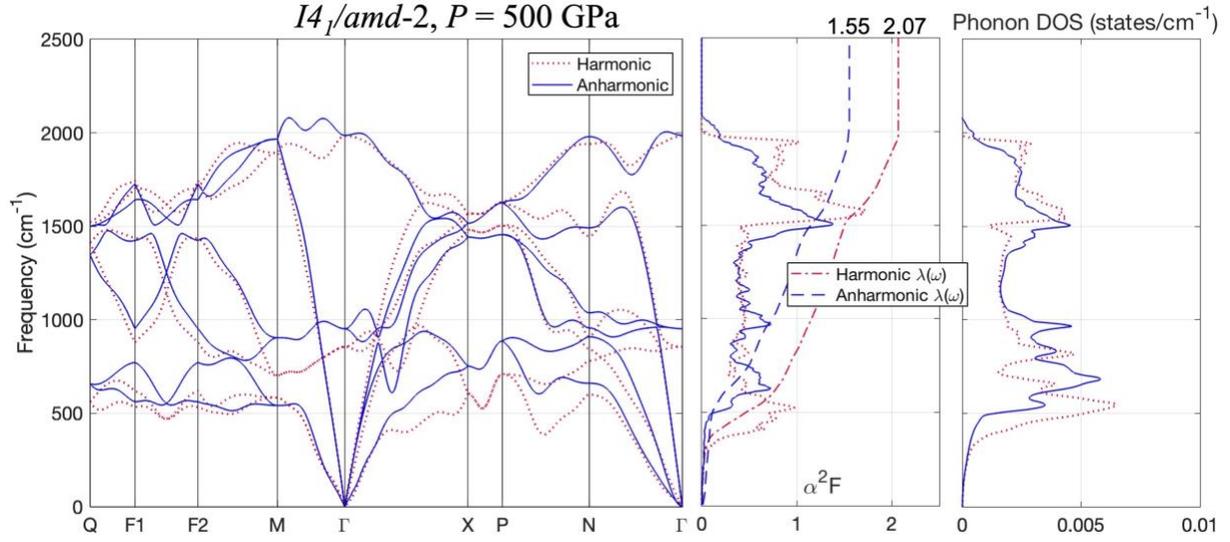

**Figure S16. Phonons and electron–phonon coupling for *I4₁/amd*-2 at 500 GPa.** The phonon dispersion relations for the *I4$_1$/amd*-2 phase of deuterium at 500 GPa pressure (left panel). The harmonic and anharmonic calculations are shown by red dashed lines and blue solid lines, respectively. The Eliashberg function $\alpha^2 F$ and the electron–phonon coupling parameter $\lambda(\omega)$ (middle panel), and the phonon densities of states (right panel).